\documentclass[%
 reprint,
 amsmath,amssymb,
 aps,
]{revtex4-2}

\usepackage{graphicx}
\usepackage{dcolumn}
\usepackage{bm}
\usepackage{hyperref}

\begin{document}

\preprint{APS/123-QED}

\title{Design of acoustic diffraction plates for manipulating ultrasound in liquid Helium}

\author{Ayanesh Maiti}
 \altaffiliation{Undergraduate Department, Indian Institute of Science, \\Bangalore - 560012, India}
\author{Dillip K Pradhan}
\author{Ambarish Ghosh}
 \email{ambarish@iisc.ac.in}
\affiliation{Department of Physics, Indian Institute of Science, Bangalore - 560012, India}

\begin{abstract}
Many experiments in liquid Helium, such as the optical imaging of exploding electron and ion bubbles, which enables research on individual particles under applied conditions, involve the usage of ultrasound generated by piezoelectric transducers. Previous studies either use planar transducers, which limits the maximum sound intensity and the spatial resolution, or curved transducers, which only allow observations at fixed foci and make it difficult to apply uniform electric fields. In this paper, we introduce the usage of acoustic diffraction plates in liquid Helium to amplify ultrasonic pressure oscillations at an arbitrary set of primary foci coupled with large counts of secondary foci, all of which can be freely moved around by changing the ultrasound frequency. The frequency dependence also allows us to generate controlled Faraday instabilities at the surface, which enables the generation of multi-electron bubbles with desired parameters.
\end{abstract}

\maketitle

\section{Introduction}
\begin{figure}
    \vspace{12pt}
    \centering
    \includegraphics[width=0.47\textwidth]{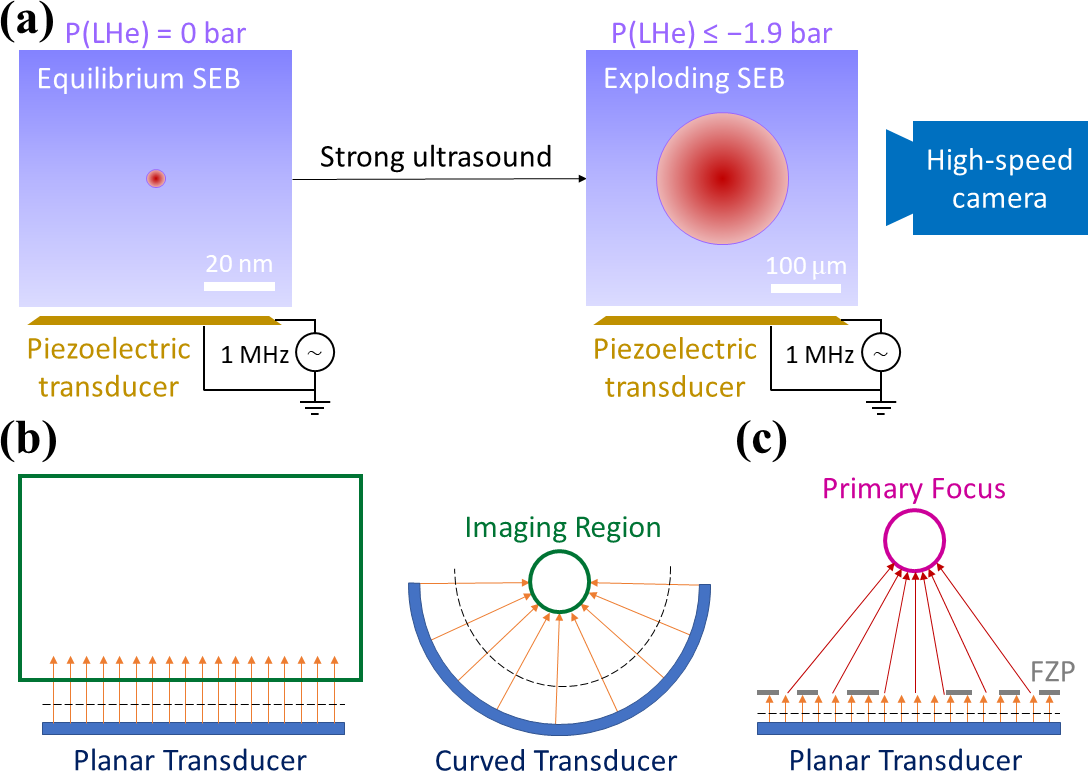}
    \caption{(a) Schematic of an SEB imaging experiment. (b) Illustration of the previously employed methods of generating strong ultrasound in liquid Helium (LHe). (c) The proposed setup for focusing ultrasound using a diffraction zone plate.}
    \label{fig1}
\end{figure}

Helium atoms repel electrons due to their fully-filled 1s$^2$ electronic configurations~\cite{elec_eion_LHe}. This fermionic repulsion causes the displacement of liquid Helium upon the approach of high-energy electrons, generating vacuum bubbles that trap the incoming electron. The resulting single electron bubbles (SEBs) possess spin relaxation and dephasing times as large as a few hours, making them a very strong candidate for quantum computing applications~\cite{QC}. This knowledge has driven many efforts to trap SEBs, manipulate their spin states, and read them out. It is thus important to develop techniques to study individual SEBs under applied electric and magnetic fields. One powerful method is to explode these SEBs using strong ultrasonic pressure oscillations, and image the exploding SEB using an optical camera, as shown in Fig.~\ref{fig1}(a). While this method is destructive in nature, it does not effect the spin state of the trapped electron, allowing the indirect measurement of the SEB spin from the observed trajectories under applied fields.

In the previous imaging studies, the ultrasound was generated using piezoelectric transducers with planar~\cite{Planar} or curved~\cite{Curved} shapes, as illustrated in Fig.~\ref{fig1}(b). While the former enables the detection of SEBs in a large region above the transducer at the cost of spatial resolution and maximum sound intensity, the latter focuses the sound into a fixed region, making us blind to any events away from their focus. Also, the usage of curved transducers makes it very difficult to introduce uniform electric fields in the experiments. Thus, we examine the utility of diffraction zone plates as focusing devices that can achieve large pressure amplification with sub-wavelength spatial resolution at movable focal points, without introducing any curvature in the system to enable the inclusion of uniform electric fields. Since these devices work on diffraction, they also generate considerable pressure amplification in many secondary regions which allows us to make observations away from the focus.

Fresnel Zone Plates (FZPs) consist of an opaque plate with transparent zones in the shape of concentric rings with well-defined radii, that focus coherent plane waves into a point on their axis of symmetry by the means of diffraction. Initially introduced in the field of optics, recent experiments have confirmed their principles to work even for acoustics for media such as air~\cite{FZP_Air} and water~\cite{FZP_Water}. In this paper, we computationally examine the application of FZPs for focusing ultrasound in liquid Helium, for an experimental setup as shown in Fig.~\ref{fig1}(c). We also attempt to further extend the concept of diffraction zone plates, in order to generate desired distribution of primary foci.

\section{Methods}
\begin{figure}
    \vspace{12pt}
    \centering
    \includegraphics[width=0.47\textwidth]{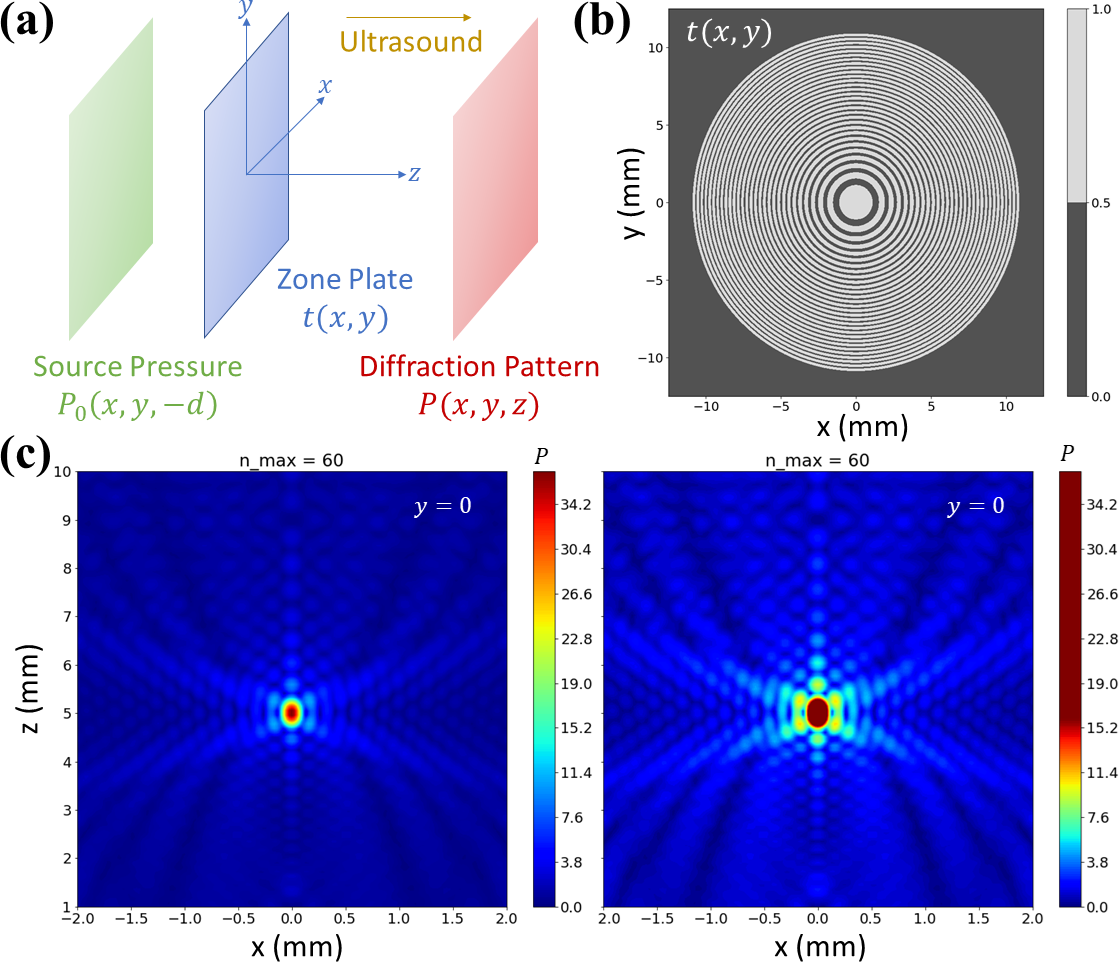}
    \caption{(a) Definition of coordinate axes illustrating the plane of transducer (green), plane of diffraction plate (blue), and plane of imaging (red). (b) Transparency profile of an FZP with $f =$~5 mm for ultrasound of frequency 1 MHz in liquid Helium, and (c) the pressure amplification it produces, with two different colour scales.}
    \label{fig2}
\end{figure}

Fig.~\ref{fig2}(a) shows our coordinate system. The diffraction plate is placed in the $xy$ plane and centered around the origin, so that the sound waves propagate along $\hat{z}$. The transducer surface is parallel to the diffraction plate and has a spacing of $d$ with the $xy$ plane in the $-z$ direction. The diffraction pattern is produced in the $z>0$ region. In previous studies, the complex pressure distribution $P(x,y,z) = A(x,y,z)\cdot e^{i\phi(x,y,z)}$ was calculated by means of a 2D real-space Fresnel integral from each source point $S\equiv(u,v,0)$ in a given plane (here $z =$ 0) to each point $I\equiv(x,y,z)$ in the imaging space~\cite{real_space_eqn}. For ultrasound of wavenumber $k$, this method involves calculating:
\begin{equation}\label{eq1}
    P(I) = \frac{k}{2\pi i}\int{P(S) \cdot f(S,I) \cdot dS}
\end{equation}
\begin{equation}\label{eq2}
    f(S,I) \equiv f\left[l(S,I),\theta(S,I)\right] = \frac{e^{ikl}}{l}\left(1-\frac{1}{ikl}\right)\cos{\theta}
\end{equation}
where $l(S,I)$ is the distance between points the $S$ and $I$, and $\theta$ is the angle between the direction of incident wave velocity (here $\hat{z}$) and the line segment $SI$. Hence we have $\cos{\theta} = z/l$ for our calculations. The pressure $P'(x,y,0)$ incident at the zone plate is calculated from the known pressure distribution generated by the transducer $P(x,y,-d)$, using equations \ref{eq1}-\ref{eq2} between the source plane $z=-d$ and the image plane $z=$ 0. The emergent pressure distribution $P(x,y,0)$, which acts as a source for the imaging space, is calculated using the known transparency profile $t(x,y)$ of the zone plate:
\begin{equation}\label{eq3}
    P(x,y,0) = t(x,y) \cdot P'[P(z = -d)](x,y,0)
\end{equation}

While this method gives accurate results as evident from the literature, we utilize an approximate method~\cite{Fourier_Acoustics} based on the technique of Fourier space optics. We start with a source plane perpendicular to the direction of propagation $\hat{z}$ at which the pressure is known (here $z =$ 0) and carry out a 2D Fourier transform $\mathcal{F}_{x,y}:(x,y) \to (k_x,k_y)$. We can then calculate the wavenumber $k_z^2 = k^2-k_x^2-k_y^2$ along the wave velocity for each Fourier component, and propagate the same to an image plane $z$, so that:
\begin{equation}\label{eq4}
    \mathcal{F}_{x,y}[P(x,y,z)] = e^{ik_zz}\cdot\mathcal{F}_{x,y}[P(x,y,0)]
\end{equation}
\begin{equation}\label{eq5}
    P(x,y,z) = \mathcal{F}^{-1}_{x,y}\left[e^{ik_zz}\cdot\mathcal{F}_{x,y}[P(x,y,0)]\right]
\end{equation}

Our method is equivalent to a far-field approximation of equations \ref{eq1}-\ref{eq2} above. Again, we use equation \ref{eq3} to move from transducer to diffraction plate to image planes. We have contrasted this method with the Fresnel integrals and found a difference of less than 1\% in the results, justifying our use of this faster algorithm. All of our codes have also been verified against the previous experimental date on FZPs for acoustics in air and water.

For our calculations we have taken the source pressure $P(x,y,-d)$ to be 1 throughout the transducer and 0 everywhere else, with a circular transducer of radius 12 mm. The diffraction plate is taken to have a transparency profile $t$ which is optimized to produce maximum amplification at the selected primary foci. For an FZP, we directly use the analytical result so that $t(r(x,y))$ alternates between 0 and 1 at discrete values of radii $r_n = n\lambda f + (n\lambda/2)^2$ for $n =1,2,...,n_{max}$ determined by the wavelength $\lambda$ of the ultrasound. This pattern is known to produce a single primary maxima at $r = 0$ and $z = f$. In this study we use a focal length $f =$ 5 mm and 1 MHz ultrasound to calculate the transparency profile of our FZP. We take the speed of sound in liquid Helium to be $c =$ 234 m/s.

\section{Results and Discussion}
Fig.~\ref{fig2}(b) shows an FZP with $n_{max} = 60$, i.e, a total of 30 transparent zones. Since the pattern is radially symmetric, the diffraction pattern would have cylindrical symmetry. So we only need to calculate $P(x,y,z)$ in the $xz$ plane to get the full pressure distribution. Fig.~\ref{fig2}(c) shows the result of this calculation in two different colour scales. The colour plot in the left side demonstrates the focusing strength of the FZP with the colour scale extending to the maximum pressure amplification of around 36$\times$. In contrast, the plot in the right shows the large number of secondary maxima by setting the final colour at 16$\times$ pressure amplification. We can see that the secondary maxima, which have a smaller spatial extent in comparison to the primary maxima, are well distributed all around the primary focus. Also, we can get around 8$\times$ pressure amplification in these secondary maxima. Thus, we already confirm that the spatial resolution of our imaging experiments is improved by the diffraction-based focusing, without introducing any curvature in our system, and also retaining the possibility of observing events away from the focus by simply increasing the supply voltage.

\begin{figure}
    \vspace{12pt}
    \centering
    \includegraphics[width=0.47\textwidth]{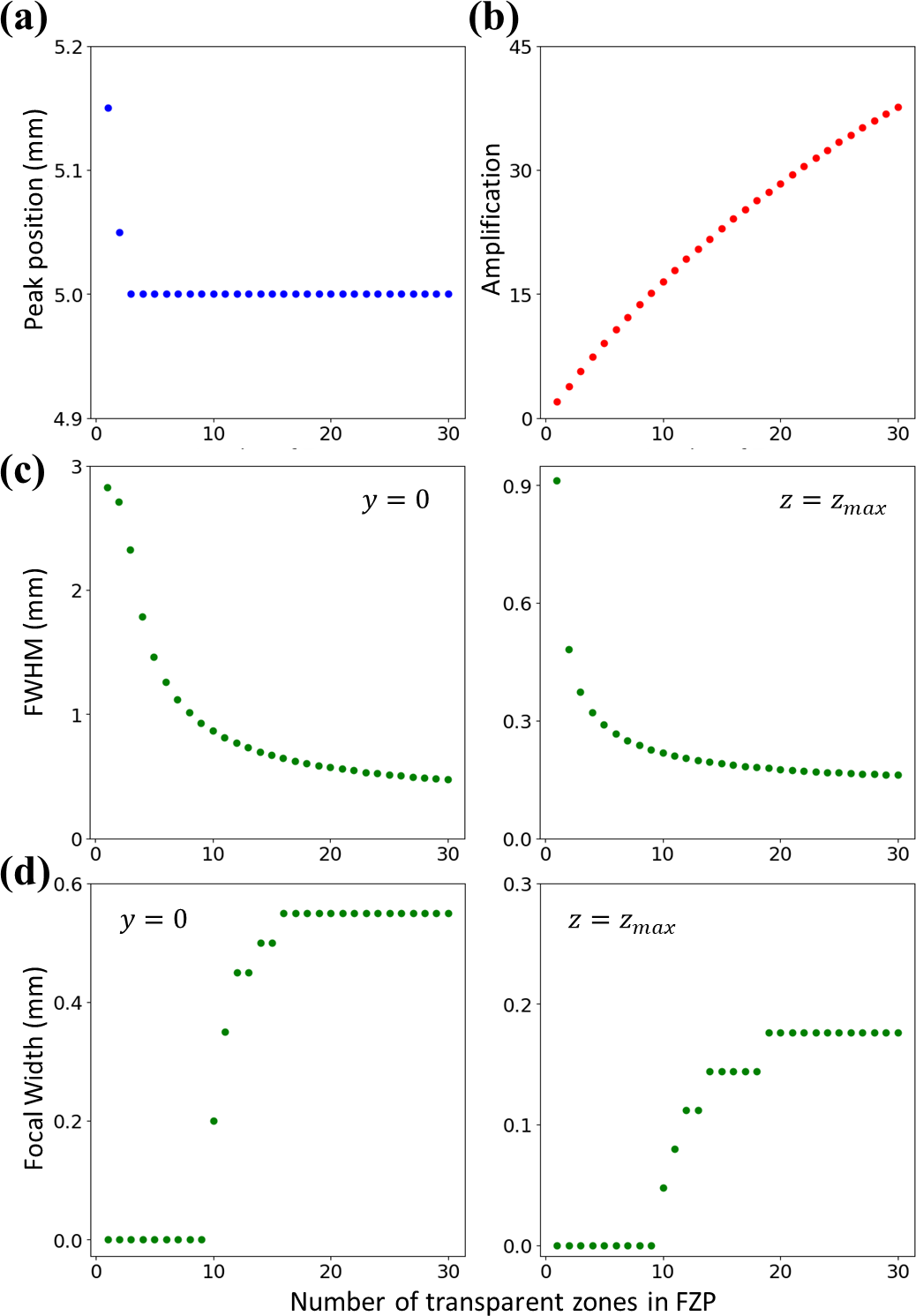}
    \caption{(a) Position of global pressure maxima and (b) the pressure amplification achieved by an FZP with $f=$~5 mm at frequency 1 MHz. (c) Full-width at half-maximum (FWHM) pressure amplification and (d) Extent of region with pressure amplification greater than 16$\times$, in the focal planes $y=0$ and $z=z_{max}$.}
    \label{fig3}
\end{figure}

\begin{figure}
    \vspace{12pt}
    \centering
    \includegraphics[width=0.47\textwidth]{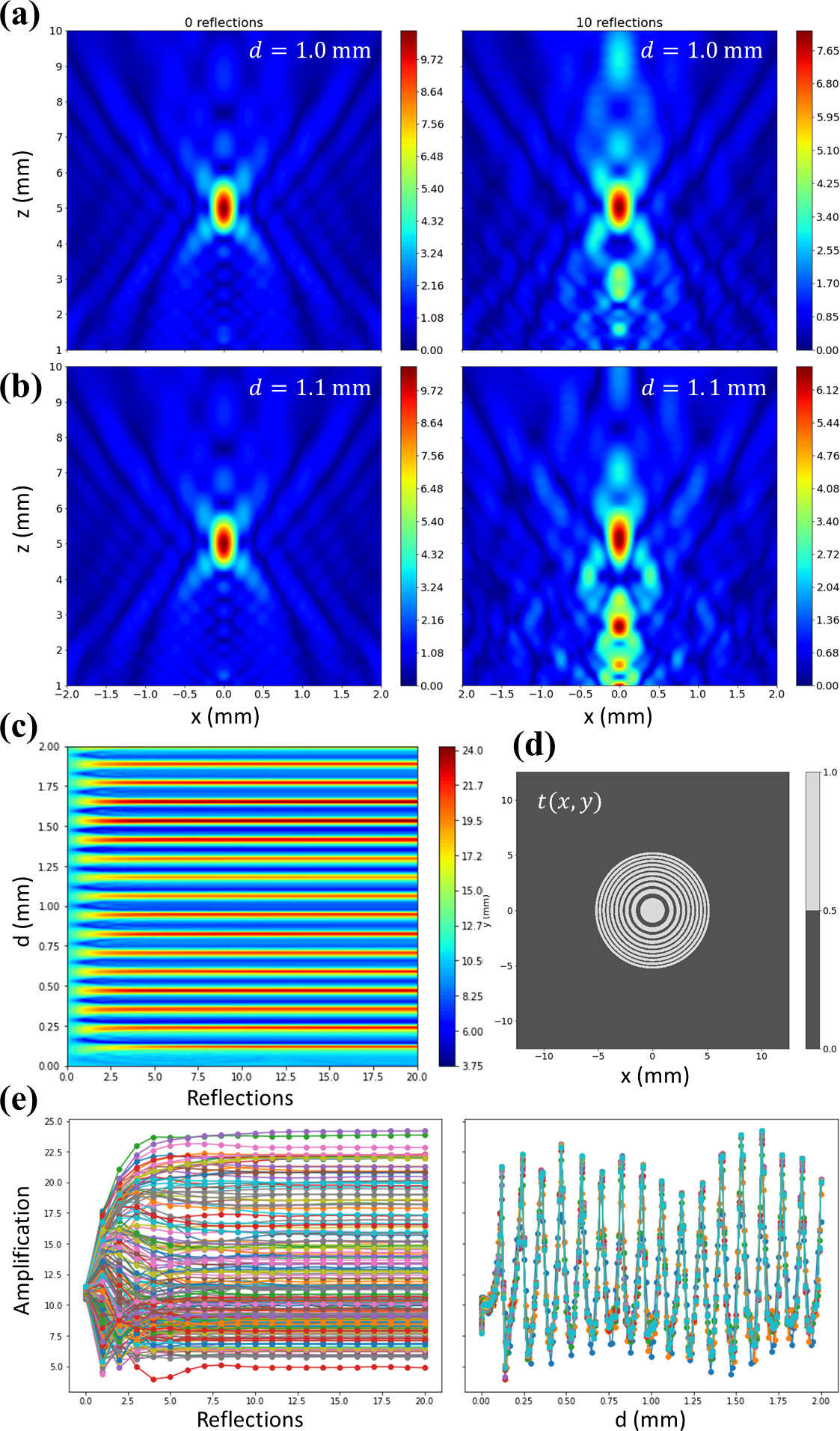}
    \caption{Pressure distribution for $d =$ (a) 1.0 mm, and (b) 1.1 mm with no reflection (left), and after 10 reflections (right). (c) Focal pressure amplification as a function of $d$ and the number of reflections for (d) an FZP of radius 4 mm designed to focus 1 MHz ultrasound at $f =$ 5 mm. (e) Focal pressure amplification as a function of the number of reflections for various values of $d$ (left) and as a function of $d$ after different numbers of reflections (right). The latter only shows the data for 5 or more reflections.}
    \label{fig4}
\end{figure}

\begin{figure}
    \vspace{12pt}
    \centering
    \includegraphics[width=0.47\textwidth]{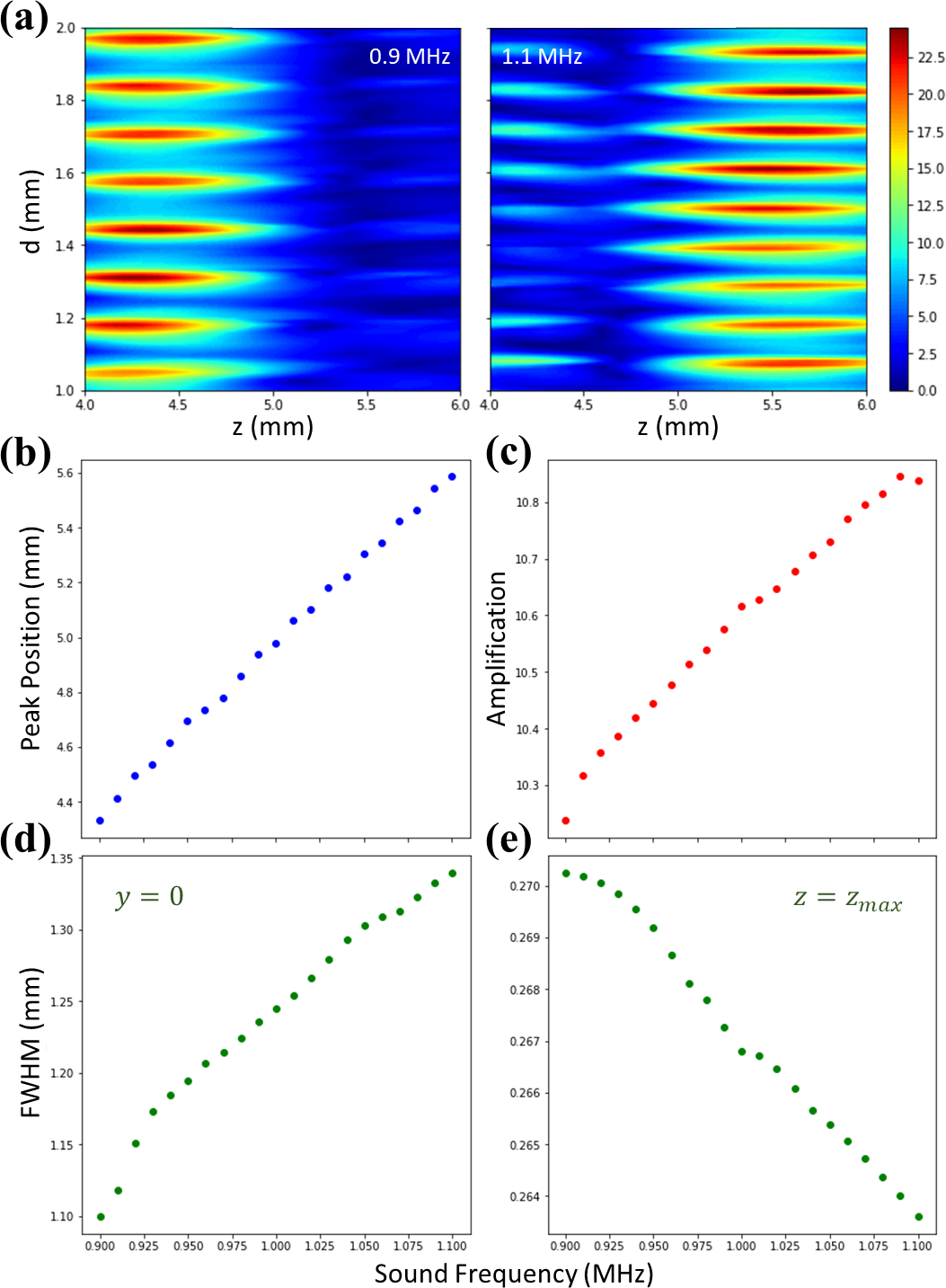}
    \caption{(a) Shifting of the peak position as a function of the sound frequency, for different values of the spacing $d$, shown for 0.9 MHz and 1.1 MHz ultrasound. (b) Position of the global pressure maxima and (c) corresponding amplification, as well as the FWHM of this primary focal spot in focal planes (d) $y =$ 0 and (e) $z=z_{max}$, as a function of sound frequency.}
    \label{fig5}
\end{figure}

\begin{figure}
    \vspace{12pt}
    \centering
    \includegraphics[width=0.47\textwidth]{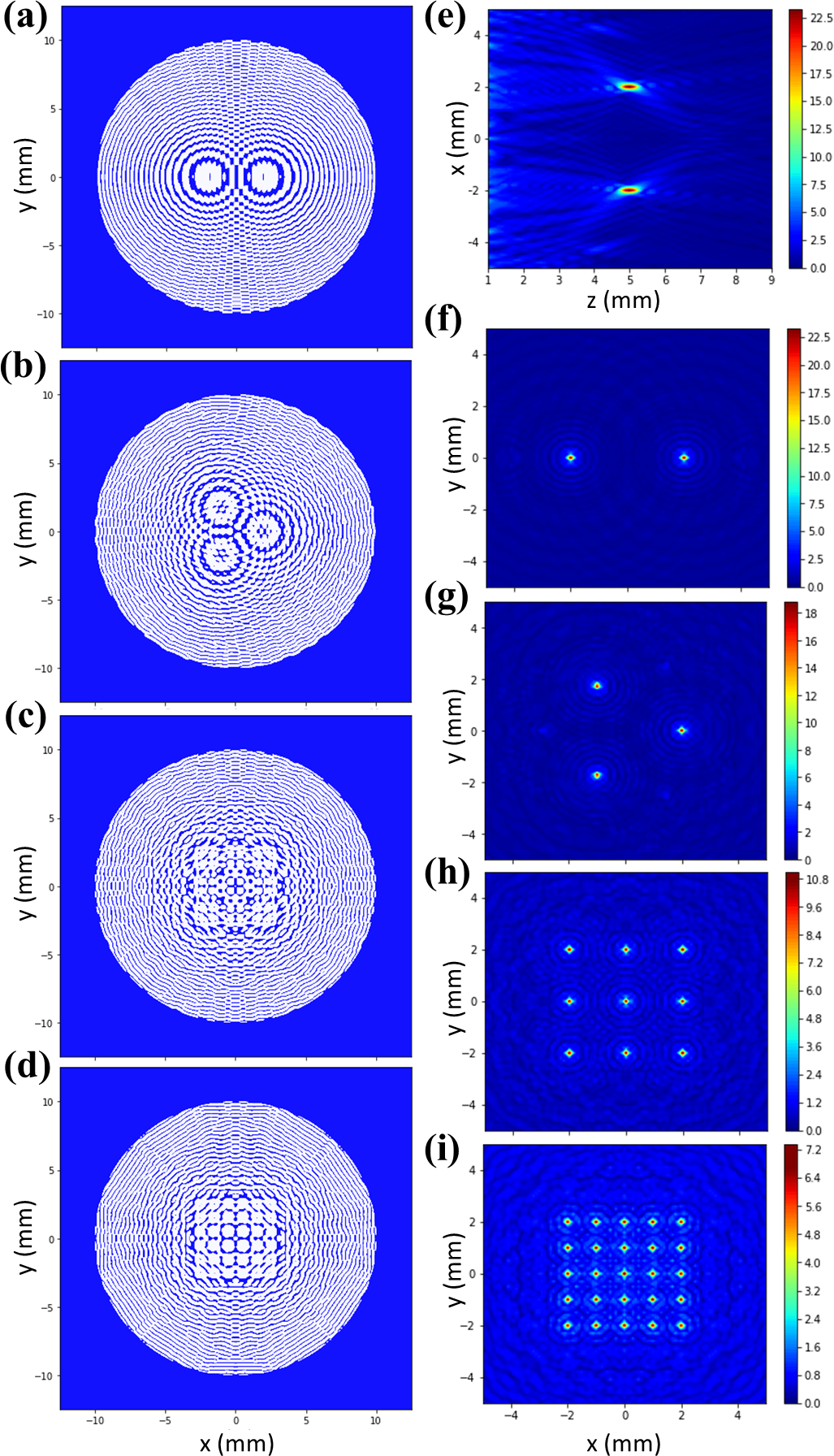}
    \caption{Diffraction zone plates designed to produce (a) two maxima, (b) three maxima at the vertices of an equilateral triangle, (c) a 3$\times$3 lattice of maxima and (d) a 5$\times$5 lattice of maxima. Plot (e) shows the pressure distribution generated by profile (a) in the $y=$ 0 focal plane. Plots (f)-(i) show the pressure distributions generated in the $z=f$ focal plane by diffraction plates of transparency profiles (a)-(d) respectively.}
    \label{fig6}
\end{figure}

Fig.~\ref{fig3}(a) shows the position $(0,0,z)$ where maximum pressure amplification is generated by an FZP, as a function of the number of transparent zones $n_{max}/2$, and Fig.~\ref{fig3}(b) shows the pressure amplification obtained at this global maximum. The former confirms that the peak is obtained at the intended location $z=f$ for zone plates with three or more zones, and the latter shows that the amplification increases as more sound is incident, both of which are expected results. Fig.~\ref{fig3}(c)-(d) shows the FWHM and spatial extent of the primary focal maximum in the $y=0$ abd $z=z_{max}$ focal planes. In all cases, we see that the spot saturates to a diffraction limiting size between 200-600 $\mu$m, which is of the order of the ultrasound wavelength of 234 $\mu$m. This means that we can get close to the best possible spatial resolution with as less as 10 transparent zones in our FZP, making it much simpler to fabricate a diffraction plate and include it in SEB imaging experiments. Also, from Fig.~\ref{fig2}(b) we see that the zones closer to the center are wider than those away from the center, so that the resolution required to fabricate an FZP is lower for smaller number of zones. Hence we can make the FZPs cheaper by trading off the maximum amplification, as per the experimental requirements.

In all of the above calculations, we have neglected any reflections between the FZP and the transducer. While our previous calculations hold strong for short pulsed measurements, in longer measurements we need to consider the interference of the reflections with the source patterns. This was not important for the previous experiments in air and water since the transducer was kept far away from the FZP to ensure plane wave incidence. This was necessary since the larger speed of sound, and hence the larger wavelength, forced the FZP to be much larger than the transducers. In our case the small wavelength allows us to place the transducer close to the FZP which enables us to maximize the incident pressure amplitude, but makes the effect of reflections more significant. Fig.~\ref{fig4}(a)-(b) demonstrates this effect considering perfect reflection at all surfaces, and shows that it strongly depends upon the spacing between the transducer and the FZP. For this calculation we have taken an FZP with $n_{max} = 10$, as shown in Fig.~\ref{fig4}(d), which we have already fabricated for our SEB imaging experiments.

From Fig.~\ref{fig4}(a) and (b), the most prominent effect change is observed in the spatial distribution of secondary maxima away from the focus, as well as a distortion and amplification change in these regions. The shape of the primary maxima is not affected as strongly, but its amplification changes significantly. Fig.~\ref{fig4}(c) shows the variation of the focal amplification as a function of the spacing $d$, and the number of reflections. We see that the amplification can be changed from its initial value of around $10\times$ to over $25\times$ or less than $4\times$. The horizontal and vertical cuts of this data, which are plotted in Fig.~\ref{fig4}(e), show that the amplification quickly saturates to its equilibrium value after 5-8 reflections, which oscillates between 0.4-2.5$\times$ the initial amplification depending on the spacing $d$ between the FZP and transducer. We observe a maximum whenever $2d = n\lambda$, which is exactly the amplification condition of a resonant cavity of size $d$ for sound waves of wavelength $\lambda$. Thus the reflections between the FZP and transducer in the region around the primary focus is effectively equivalent to coupling a resonant cavity before the diffraction-based focusing.

Now that we have accounted for all major effects in our calculations, we study the changes in the diffraction pattern when we modify the ultrasound frequency by controlling the AC supply to the piezoelectric transducer. As shown in Fig.~\ref{fig5}(a), the primary focus shifts on changing the frequency. Fig.~\ref{fig5}(b)-(e) shows the shift in the focal position, the change in the maximum amplification, and the corresponding changes in the FWHM in the $y=0$ and $z=z_{max}$ focal planes respectively. These calculations do not take the reflections into account, since we can decouple their effects near the focus. All parameters are found to depend linearly on the sound frequency, making it easy to control the pressure distribution during an experiment. With just a 10\% change in frequency, we can move the focus by over 500 $\mu$m, which is more than twice the sound wavelength! Also, the amplification and radial FWHM change just around 2\% with this, meaning that we are essentially moving the same primary region along the $z$-axis by making controlled adjustments in the supply frequency.

While the previous analysis already demonstrates the utility of an FZP to focus ultrasound in liquid Helium, diffraction plates can be made with different goals than to focus all incident sound to a single point. Fig.~\ref{fig6}(a)-(d) demonstrate some of the simplest extension one can make. In order, we have solved the optimization problem for desired pressure distributions consisting of two identical maxima, three maxima arranged in an equilateral triangle, a 3$\times$3 lattice of maxima and a 5$\times$5 lattice of maxima, all in the same plane perpendicular to the direction of sound propagation. The co-planarity is employed so that we can simultaneously observe events at the various maxima, even for short-pulsed measurements. Fig.~\ref{fig6}(e) and (f) show the diffraction pattern generated by a diffraction plate with transparency profile as shown in Fig.~\ref{fig6}(a), while Fig.~\ref{fig6}(g)-(i) show the pressure distribution generated by transparency profiles as shown in Fig.~\ref{fig6}(b)-(d) respectively. In all cases we are able to closely achieve the intended patterns, demonstrating the flexibility of this technique.
\\

\section{Conclusions}
In conclusion, our calculations show that diffraction plates can be zero-curvature focusing device to generate strong pressure oscillations in liquid Helium, which allow the flexibility to move the primary focus and observe events away from the focus in addition to improving the spatial resolution to sub-wavelength length scales and the ultrasound intensity by over 1200$\times$. The focal intensity can also be further enhanced to over 7500$\times$ by adjusting the size of the resonant cavity between the diffraction plate and the ultrasound transducer. This comes with the additional capability to directly control the focal position by means of the ultrasound frequency, and designing customized transparency profiles to generate desired pressure distributions!

\bibliography{ref}

\end{document}